\documentclass[sigconf,prologue,table,balance=false]{acmart}

\acmConference[ICSE 2024]{46th International Conference on Software Engineering}{April 2024}{Lisbon, Portugal}
\AtBeginDocument{%
  \providecommand\BibTeX{{%
    \normalfont B\kern-0.5em{\scshape i\kern-0.25em b}\kern-0.8em\TeX}}}

\usepackage{amsmath,amsfonts}
\usepackage{algorithmic}
\usepackage{graphicx}
\usepackage{textcomp}
\usepackage{xcolor}
\def\BibTeX{{\rm B\kern-.05em{\sc i\kern-.025em b}\kern-.08em
 T\kern-.1667em\lower.7ex\hbox{E}\kern-.125emX}}

\usepackage[framemethod=tikz]{mdframed} % 
\usepackage{enumitem}
\usepackage{soul}
\usepackage{stfloats}

\usepackage{soul}
\usepackage{colortbl}
\usepackage{adjustbox}
\usepackage{caption}
\usepackage{booktabs} 
\usepackage{multirow}
\usepackage{tikz}
\usepackage{xcolor}
\usepackage{refcount}
\usepackage{hyperref}
\usepackage{tabularx}
\usepackage{threeparttable}
\usepackage{multicol}
\usepackage{float}

\usepackage[most]{tcolorbox}

\definecolor{purplish}{HTML}{D8D0E3}
\definecolor{purplishlight}{HTML}{EBE7F1}
\definecolor{purplishdark}{HTML}{20aee5}

\newtcolorbox[auto counter,number within=section]{rqbox}[2]{
    nameref=#1,
    title=\small{#1}, 
    enhanced,
    attach boxed title to top left={yshift=-6pt, xshift=8pt},
    boxed title style={size=small,boxsep=1pt},
    colframe=purplishdark,colback=white,colbacktitle=purplishdark,
    boxsep=2pt,left=2pt,right=2pt,top=6pt,bottom=2pt,middle=2pt
}

\begin{document}

\title{Applying Large Language Models API to Issue
Classification Problem}

\author{Gabriel Aracena}
%\authornote{All authors contributed equally to this research.}
\email{GTeixeiraL@my.gcu.edu}
\affiliation{%
  \institution{Grand Canyon University}
  \streetaddress{3300 W Camelback Rd}
  \city{Phoenix}
  \state{Arizona}
  \country{USA}
  \postcode{85017}
}

\author{Kyle Luster}
%\authornotemark[1]
\email{kluster1@my.gcu.edu}
\affiliation{%
  \institution{Grand Canyon University}
  \streetaddress{3300 W Camelback Rd}
  \city{Phoenix}
  \state{Arizona}
  \country{USA}
  \postcode{85017}
}

\author{Fabio Santos}
\email{fabiomarcos.deabre@gcu.edu}

\affiliation{%
  \institution{Grand Canyon University}
  \streetaddress{3300 W Camelback Rd}
  \city{Phoenix}
  \state{Arizona}
  \country{USA}}

\author{Igor Steinmacher}
\email{igor.steinmacher@nau.edu}

\affiliation{%
  \institution{Northern Arizona University}
  \city{Flagstaff}
  \state{Arizona}
  \country{USA}
}

\author{Marco A. Gerosa}
\email{marco.gerosa@nau.edu}

\affiliation{%
 \institution{Northern Arizona University}
 \streetaddress{1900 S Knoles Dr.}
 \city{Flagstaff}
 \state{Arizona}
 \country{USA}}

%%
%% By default, the full list of authors will be used in the page
%% headers. Often, this list is too long, and will overlap
%% other information printed in the page headers. This command allows
%% the author to define a more concise list
%% of authors' names for this purpose.
\renewcommand{\shortauthors}{Aracena and Luster, et al.}

\begin{abstract}
Effective prioritization of issue reports is crucial in software engineering to optimize resource allocation and address critical problems promptly. However, the manual classification of issue reports for prioritization is laborious and lacks scalability. Alternatively, many open source software (OSS) projects employ automated processes for this task, albeit relying on substantial datasets for adequate training.
This research seeks to devise an automated approach that ensures reliability in issue prioritization, even when trained on smaller datasets. Our proposed methodology harnesses the power of Generative Pre-trained Transformers (GPT), recognizing their potential to efficiently handle this task. By leveraging the capabilities of such models, we aim to develop a robust system for prioritizing issue reports accurately, mitigating the necessity for extensive training data while maintaining reliability.
In our research, we have developed a reliable GPT-based approach to accurately label and prioritize issue reports with a reduced training dataset. By reducing reliance on massive data requirements and focusing on few-shot fine-tuning, our methodology offers a more accessible and efficient solution for issue prioritization in software engineering. Our model predicted issue types in individual projects up to 93.2\% in precision, 95\% in recall, and 89.3\% in F1-score.
\end{abstract}

%
%\begin{CCSXML}
%<ccs2012>
%<concept>
%<concept_id>10003120.10003121.10011748</concept_id>
%<concept_desc>Human-centered computing~Empirical studies in HCI</concept_desc>
%<concept_significance>500</concept_significance>
%</concept>
%</ccs2012>
%\end{CCSXML}
%
%\ccsdesc[500]{Human-centered computing~Empirical studies in HCI}
%%%
%% Keywords. The author(s) should pick words that accurately describe
%% the work being presented. Separate the keywords with commas.

\maketitle

\section{Introduction}

%Becoming a contributor in an Open Source Software (OSS) project is challenging. Newcomers face several barriers \cite{10.1145/2675133.2675215} when onboarding, and many give up \cite{8453084,EleniRetention,BogdanDisengagement}. With so many newcomer dropouts, there is a wasted opportunity to develop the technology workforce, recruit and retain contributors to projects \cite{steinmacher2013newcomers}.

%Previous work~\cite{wang2011bug,steinmacher2015understanding,steinmacher2015systematic,stanik2018simple} reveals that newcomers particularly struggle to find an appropriate task. 
The onboarding of newcomers is important to keep OSS projects sustainable~\cite{steinmacher2015systematic}. One of the initial steps of the onboarding process in an OSS project is to find an appropriate task (e.g. bugs, features, etc) to work with. The literature shows that this is a crucial step to determine the future of the newcomer in the project~\cite{wang2011bug,steinmacher2015understanding}.  
One strategy adopted by the communities is to add labels to the issues to help new contributors find the most appropriate ones~\cite{steinmacher2018let,santos2022how}. However, labeling issues in big projects is time-consuming and demands efforts from the (already overloaded) maintainers~\cite{9057411}. 

More recently, many researchers proposed approaches to label issue types automatically to help managers prioritize and allocate better available resources. \citet{kallis2019tickettagger, kallis2020tickettagger} use \textit{fastText} to classify issues as \textit{bug}, \textit{feature} or \textit{question}. Still, \citet{colavito2023few} employed SETFIT in the last NLBSE competition to predict issue types. \citet{santos2021can} predicted skills to solve an issue using API domains as a proxy. The work was extended to use Social Network Analysis (SNA), improving the predictions \cite{santos2023tell}. Finally, a tool is available to OSS communities to recommend issues based on skills informed by developers \cite{DBLP:conf/msr/VargovichSPGS23}. 

In this study, %tailored to the tool competition, 
we leverage OpenAI API to create a fine-tuned model to classify issues as bugs, features, or questions. We reached an F1-score of 82.8\%. OpenAI is the innovative company behind the development of ChatGPT---the most notorious of the many Large Language Models (LLM) they have built. 
An LLM is a sophisticated neural network model that undergoes training using extensive datasets, including books, code, articles, and websites. This training enables the model to grasp the inherent patterns and relationships within the language for which it was trained. Consequently, the LLM can produce cohesive content, such as grammatically accurate sentences and paragraphs, replicating human language, or syntactically precise code snippets \cite{ozkaya2023application}
and it is capable of adapting incredibly when fine-tuned. %(describe high-level what is behind it: deep learning? ).%

\textbf{RQ1: To what extent can we predict the issue types using OpenAI's fine-tuning API?} To answer RQ1, we 
fine-tuned the gpt-3.5-turbo base model provided in the OpenAI API. Fine-tuning is the process of giving specific and niched training for a pre-trained model. Fine-tuning through the OpenAI API is a multi-step process that involves simulating conversations with the LLM and telling them the expected response. We used the title and body of the issues in our training data as part of our prompt and the correct label as the expected response.  

%approach to predict the API-domain labels. We also explored the influence of task elements (i.e., title, body, and comments) and machine learning setup (i.e., n-grams and different algorithms) on the prediction model. 
Overall, we found that by pre-processing the issue title and body, we can predict the issue types with a macro average of 83.24\% precision, 82.87\% recall, and 82.8\% F-measure. This yield barely surpassed the baseline reported in the competition~\cite{nlbse2024}. 

%because projects require skills in different languages, frameworks, databases, and Application Programming Interfaces (APIs). 

%APIs usually encapsulate modules that have specific purposes (e.g., cryptography, database access, logging, etc.), abstracting the underlying implementation. If the contributors know which types of APIs will be required to work on each issue, they could choose tasks that better match their skills or involve skills they want to learn. 

%domains of APIs to facilitate contributors' task selection. Since an issue may require knowledge in multiple APIs, we applied a multi-label classification approach, which has been used in software engineering for purposes such as, classifying questions in Stack Overflow (e.g., \citet{xia2013tag}) and detecting types of failures (e.g., \citet{feng2018empirical}) and code smells (e.g., \citet{guggulothu2020code}). 

%By employing an exploratory case study and a user study, we aimed to answer the following research questions:

%\section{Related Work}
\label{sec:related}

%Previous research addressed the issue classification problem by employing many different techniques. \citet{kallis2020tickettagger}, employed $J48$ ML algorithm to predict bug, enhancement, and questions and obtained 0.83 for precision, 0.82 for recall, and 0.83 for F-measure. \citet{el2020automatic} addressed the misclassification problem for inter and intra-project situations, classifying issues as bug/non-bug. 

%Large language models have been used to architect a service-based software \cite{ahmad2023towards} with the support of bots. Documentation is also a prominent area where LLMs may be employed \cite{ozkaya2023application}. \citet{nathalia2023artificial} evaluated the performance of code automation using LLMs.

\section{Method}

%(describe without code details the process to process the dataset and obtain the results. Divide it into subsections. Example below. May include others or change the name. Get ideas from here \url{https://ieeexplore.ieee.org/stamp/stamp.jsp?tp=&arnumber=10189133} and here: \url{https://ieeexplore.ieee.org/stamp/stamp.jsp?tp=&arnumber=10189129}.% 

\begin{figure}[!hbt]
\centering
\includegraphics[width=.45\textwidth]{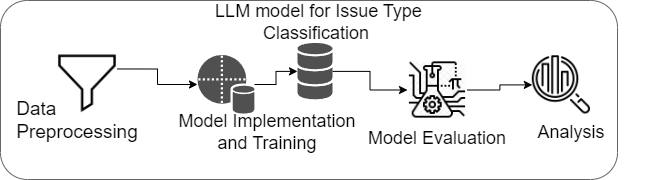}
\caption{Method}
\label{fig:method}
\end{figure}

Figure \ref{fig:method} depicts our method, composed of the data preprocessing, model implementation and training, model evaluation, and analysis phases. 

\subsection{Data Preprocessing and Cleaning}

\subsubsection{Introduction to the Dataset}

The dataset for the \textit{NLBSE'24 Tool Competition on Issue Report Classification} is a collection of 3,000 labeled issue reports, which have been extracted from five open-source projects. The data was collected over 21 months, from January 2022 to September 2023, ensuring a wide range of issues and scenarios are covered.

\subsubsection{Attributes of Each Issue Report}
Each issue report in the dataset is characterized by four key attributes:

\begin{enumerate}
\item \textbf{Repository}: The name of the open-source project from which the issue report was extracted.
\item \textbf{Label}: The category that the issue report falls into.
\item \textbf{Title}: The title of the issue report.
\item \textbf{Body}: The main content of the issue report.
\end{enumerate}

\subsubsection{Labeling and Exclusions}
Each issue has only one label; labeled either bug, feature, or question.

\subsubsection{Noise Removal and Data Preparation}
\begin{itemize}
  \item Method 1: The method involves the removal of double quotation marks, elimination of specific string patterns, lowercase conversion for standardization, and removal of emojis, URLs, HTML tags, special characters, and punctuation. The method also addressed consecutive whitespace and restricted word length to 20 characters, as any string longer than that is not likely to be a word but noise.
  \item Method 2: The second method mirrors Method 1's cleaning process with a handful of appropriate modifications. It eliminates more string patterns, modifies handling URLs and HTML tags by substituting them with universal identifiers (<URL> and <HTML\_TAG>), and replaces usernames and image links with their own identifiers, <USER> and <IMAGE>, respectively. Lastly, it strips Markdown syntax from the text, striving to preserve only the tokens necessary for analysis.
\end{itemize}

\subsubsection{Data Segmentation and Labeling}
The CSV file was split into five repositories, each with 300 categorized issue reports ("bug," "feature," or "question"), and the data provided was prelabeled.

\subsubsection{Format Conversion for Model Input}
Repository training data frames were converted into JSON line files, structured as a conversation, and handed off the to API. Each file contains a prompt, (\texttt{user\_message}).  We decided to use the prompt, \textit{"Classify, IN ONLY 1 WORD, the following GitHub issue as 'feature', 'bug', or 'question' based on its title and body}:". The second half of the conversation is called (\texttt{assistant\_message}), and it contains the model's classification ('bug,' 'feature,' or 'question'). Both halves are concatenated and assigned to the variable (\texttt{conversation\_message}), these variables are then appended iteratively into a single JSON line file.

\subsection{Model Implementation and Training}
\subsubsection{Invoking the API}
The API invoked in our code is OpenAI's fine-tuning API. Our chosen model is the gpt-3.5-turbo model for natural language understanding and generation tasks. This API enables interaction with language models capable of understanding prompts and generating coherent and contextually relevant responses. We invoked the API using OpenAI’s Python library.

Firstly, we invoke the OpenAI API. Next, the training files are uploaded to the server. Using these files, gpt-3.5-turbo creates fine-tuned models for each repo. These models will be retrieved for future use. To test the model, the interaction with the API occurs through the \texttt{create()}\footnote{openai.chat.completions.create()} method, where a user prompt is constructed. Here we reuse the prompt from the training cycle, append the testing data, and pass it to the API. The constructed prompt returns its classification of the issue. The API invocation involves parameters such as the model to use, the maximum number of tokens to generate, and the temperature for controlling the randomness of responses---all of which have been tuned to yield optimal results. The model utilized was the fine-tuned model corresponding to its testing dataset and the maximum number of tokens to be returned was set to 1. The strings 'feature,' 'bug,' and 'label,' were all verified by the OpenAI Tokenizer (https://platform.openai.com/tokenizer) to have 1 token each. The temperature was set to 0.0 to minimize randomness in the answers, providing consistent results.

\subsubsection{Model Fine-tuning Process}
Fine-tuning a model through OpenAI's API involves a systematic multi-step procedure that requires a comprehensive grasp of the API documentation. As previously outlined, the fine-tuning process was tailored for each repository, demanding dedicated models for enhanced performance.

As explained in the "Format Conversion for Model Input" section, we created JSON line files customized for individual repositories. These files were uploaded to OpenAI's server to serve as training data for fine-tuning the models. To start the fine-tuning process, we initiated a job to refine the base model using the specified training file. Each job was queued in the cloud environment, leading us to regularly check their statuses every 30 minutes. After around 5 hours, all jobs were finished, and the fine-tuned models were prepared for use.

Upon completion, the server assigned unique identifiers for each fine-tuned model. For instance, the model tailored for the 'facebook/react' repository was assigned the ID \textit{ft:gpt-3.5-turbo-0613:gcucst440:fb-issueclassifier:8LLGMnAI}.

Initially, the default hyperparameters were used during the fine-tuning process, automatically configuring the learning rate multiplier, number of epochs, and batch size. All models began with the default number of 3 epochs; however, as we conducted our training, step metrics revealed room for improvement in certain models. Given this, each model's epoch count was tailored for optimal performance. Future improvements might involve fine-tuning other hyperparameters for optimization.

\subsection{Performance Evaluation Metrics}
%For our performance evaluation metrics, we iterated over the result of each issue on the test dataset and created two lists, one of the true labels and one of the predicted labels of our model. With those lists, we were capable of generating 5 confusion matrices, one for each repo/model, and then 1 combined confusion matrix for the overall result. By gathering the values of True Positive (TP), True Negative (TN), False Positive (FN), and False Negative (FN) we were capable of calculating precision, recall, and F1-Score of every model and a combined result.
%We employed precision, recall, and F1-score as our primary metrics, generating confusion matrices for each repository and a combined one for overall performance. This approach allowed us to evaluate the model’s accuracy and reliability in classifying issue reports across different contexts
We obtained our results by comparing the predictions with the ground truth. To do this, two lists were created, one with the correct labels from the testing dataset and one with the predicted labels from the fine-tuned model. We used precision\_score, recall\_score, and f1\_score from \textit{sklearn.metrics} %to avoid errors 
to standardize the results obtained across all repositories. Using the metrics functions from sklearn and the lists, we were capable of calculating precision, recall, and f1-score for all models.

\subsection{Getting the results}
%We generated individual CSV files containing the confusion matrix and metrics for each model. Following the evaluation of the five models, we compiled these metrics into a combined CSV file, consolidating the values for True Positives (TP), True Negatives (TN), False Positives (FP), and False Negatives (FN).
%Results from each model were compiled into individual and combined CSV files containing confusion matrices and performance metrics. This compilation provided a comprehensive view of the model's performance, highlighting areas of strength and those needing improvement
After computing metrics for all repositories, we calculated the average for each metric and each label to obtain the overall results present in table~\ref{tab:completemetrics}.

\section{Results}\label{sec:results}

To answer our RQ1, (to what extent can we predict the issue types using OpenAI's fine-tuning API?) we tested different repo datasets to their respective fine-tuned model.

%(We can report the results comparing different set ups, ex: cleaning x non-cleaning, title + body or only body, etc, if they are interesting.)%

Table~\ref{tab:completemetrics} shows the average f1-score results varied from 76.65\% to 87.08\%. This can be explained by the differences in how issues were written and the particularities of each repository. Overall, when using \textsc{title} and \textsc{body} combined we reached overall precision of 83.24\% recall of 82.87\% and F1-score of 82.80\%

\begin{table}[ht]
\centering
\caption{Detailed Issue Report Classification Complete Table: Metrics by Repo and by label}
\small
CM = Cleaning Method ; E = Total Epochs ; P = Precision ; R = Recall % Your note
\label{tab:completemetrics}
\begin{tabular}{l|l|l|l|l|l|l}
\hline
\textbf{Repo} & \textbf{CM} & \textbf{E}  & \textbf{Label} & \textbf{P} & \textbf{R} & \textbf{F1}    \\ \hline
facebook   & 1 & 3     & bug            & 0.8333             & \textbf{0.9500}            & 0.8878          \\
facebook   & 1 & 3     & feature        & 0.8557             & 0.8900            & 0.8725          \\
facebook   & 1 & 3     & question       & 0.9024             & 0.7400            & 0.8132          \\ \hline
facebook   & 1 & 3    & average        & 0.8635               & 0.8600            & 0.8579   \\
\hline
tensorflow & 2 & 10 & bug            & 0.9072             & 0.8800            & \textbf{0.8934}          \\
tensorflow & 2 & 10 & feature        & \textbf{0.9318}             & 0.8200            & 0.8723          \\
tensorflow & 2 & 10 & question      & 0.7913             & 0.9100            & 0.8465          \\ \hline
tensorflow & 2 & 10 & average& \textbf{0.8768}             & \textbf{0.8700}            & \textbf{0.8708} \\
\hline
microsoft & 1 & 6      & bug            & 0.8511             & 0.8000            & 0.8247          \\
microsoft & 1 & 6       & feature        & 0.8131                & 0.8700            & 0.8406          \\
microsoft & 1 & 6      & question       & 0.7980             & 0.7900             & 0.7938          \\ \hline
microsoft & 1 & 6      & average& 0.8207             & 0.8200             & 0.8198 \\
\hline
bitcoin & 1 & 3        & bug            & 0.7339             & 0.8000             & 0.7656          \\
bitcoin & 1 & 3        & feature        & 0.8318             & 0.8900            & 0.8599          \\
bitcoin & 1 & 3        & question       & 0.7381             & 0.6200            & 0.6739          \\ \hline
bitcoin & 1 & 3       & average        & 0.7679             & 0.7700            & 0.7665 \\
\hline
opencv & 2 & 6          & bug            & 0.7288             & 0.8600            & 0.7890          \\
opencv & 2 & 6        & feature        & 0.9091             & 0.8000             & 0.8511          \\
opencv & 2 & 6         & question       & 0.8617             & 0.8100            & 0.8351          \\ \hline
opencv & 2 & 6         & average        & 0.8332             & 0.8233            & 0.8250 \\
\hline \hline
overall   & NA & NA           & bug            & 0.8109             & 0.8580           & 0.8321          \\
overall   & NA& NA           & feature        & \textbf{0.8683}             & \textbf{0.8540}           & \textbf{0.8593}          \\
overall   & NA & NA           & question       & 0.8183             & 0.7740           & 0.7925          \\ \hline \hline
overall    & NA & NA          & average        & 0.8324             & 0.8287            & 0.8280 \\ \hline \hline
\end{tabular}
\end{table}

%(Next state the findings in terms of what is the best result and how the best compares with the others. Example below.)  

%(Example: Random Forest (RF) and Neural Network Multilayer Perceptron (MLPC) were the best models when compared to Decision Tree (DT), Logistic Regression (LR), MlKNN, and Dummy algorithms. Random Forest outperformed these four algorithms with large effect sizes considering F-measure and precision.)

%(Summarize the findings. Example below.)

\textbf{\emph{RQ1 Summary.}} It is possible to predict the GitHub Issue labels with precision of 83.24\%, recall of 82.87\%, and F-measure of 82.8\% using fine-tuned gpt-3.5-turbo base models, with \textsc{title} and \textsc{body} as features.
\footnote{ Check the full repository at \url{https://github.com/G4BE-334/NLBSE-issue-report-classification}}
\section{Discussion}

%(Discuss relevant topics. Example. Why the predictions have different results regarding the issue types? Any reason for this?  Example below:)
%(Example: \noindent\textbf{What are the effects of the characteristics of the data corpus? } Observing the results reported for different corpora used as input, we noticed that the baseline model created using only the issue body had similar performance to the models using the issue title, body, and comments or better performance than the model using only title. By inspecting the results, we noticed that by adding more words to create the model, the matrix of features becomes sparse and does not improve the classifier performance.
Observing the complete results (Table~\ref{tab:completemetrics}) something fascinating caught our eyes: why are the metrics so different regarding both the repository and the label? The issues labeled as questions were by far the worst on (almost) every repository. Why is that? It seems that question is a bad label name for GitHub issues. It seems to be related to the fact that question is just a very broad label that consequently causes users to label issues as a question wrongfully, and they could be better labeled as something else. When analyzing the data and checking the question label issues, it is very hard even to define a question label issue when reading the title and body. When compared to feature and bug, it is clear that question issues are not very well defined.

Similarly, when comparing the metrics on different repositories it is clear that the results achieved for the bitcoin repository were worse when compared to the other repos. When we compare the F1-score with the tensorflow repo (87.08\%) the F1-score of the bitcoin repo is more than 10\% worse (76.65\%). Why are the results so different across different repositories? After further analysis, we concluded that it is due to bad labeling and description of the issues by the developers who work on each repository. We were able to conclude that by comparing our results with the baseline metrics and saw that the baseline model had the same issue, so it is clear to us that the model performed poorly because of the data. Additionally, we saw that many of the Bitcoin issues were written without much clarity, solidifying our hypothesis. 

\noindent\textbf{What are the difficulties in labeling?} Each repository has specific concepts, technologies, and domain-related topics that vary from one repo to another. Moreover, the issue description style is also different, making automated labeling approaches more challenging.
%As previously mentioned, the definition of each label type is one of the reasons why models might perform poorly. Regarding OSS it might be hard for a newcomer to label the issue they are opening properly, and sometimes they might even label one thing but describe it as another due to inexperience. 

%(We can show the confusion metric if this helps. Example below)

%(Example: Looking at Table~\ref{tab:confusionM} and comparing it with the aforementioned co-occurrence data, we can determine some expectations and induce some predictions. For example, the ``database" label occurred more frequently when we had ``UI" and ``IO". So, it is possible to guess when an issue has both labels, and we likely can suggest a ``database ``label, even when the machine learning algorithm could not predict it. The same can happen with the ``Latex" label, which co-occurred with ``IO" and ``Network". A possible future work can combine the machine learning algorithm proposed in this work with frequent itemset mining techniques, such as apriori \cite{Apriori}. Thus, we can find an association rule between the previously classified labels.)

Looking at the confusion matrix below in Table \ref{tab:confusionM}, we can see %also infer the results obtained and understand that indeed our metrics were right. The 
models have varying degrees of success in classifying different types of issues (bugs, features, questions) across the different repositories. The model performs consistently in identifying 'bug' labels across repositories, with TP ranging from 80 to 89, suggesting that the features used by the model are good indicators of this class. The models seem to perform well on 'feature' classification, with relatively high TP and higher FP than 'bug' classification. This could indicate confusion between 'feature' and other types of issues, leading to more false alarms. There is a notable variance in the model's ability to correctly classify 'question' labels, with the lowest TP observed in the 'bitcoin/bitcoin' repository.

The 'facebook/react' repository has relatively balanced classification across all labels, indicating that the model may have learned distinctive features of each label well in this context. The 'tensorflow/tensorflow' repository has the highest FP for 'question' classification, suggesting the potential over-classification of this label. The 'microsoft/vscode' repository tends to miss 'feature' and 'question' issues (as indicated by higher FN).

%\todo{R2 QIIIC} \draft{new table VII after run the SMOTE}

\begin{table}[h!]
 \begin{center}
 \caption{Confusion Matrix for all projects and labels}
 \label{tab:confusionM}
\begin{tabular}{l|l|l|l|l|l}
 \hline
\textbf{Repository}   & \textbf{Label} & \textbf{TP} & \textbf{FP} & \textbf{FN} & \textbf{TN} \\ \hline
facebook/react        & bug            & 89          & 15          & 11          & 185         \\
facebook/react        & feature        & 95          & 19          & 5           & 181         \\
facebook/react        & question       & 74          & 8           & 26          & 192         \\ \hline
tensorflow/tensorflow & bug            & 82          & 6           & 18          & 194         \\
tensorflow/tensorflow & feature        & 88          & 9           & 12          & 191         \\
tensorflow/tensorflow & question       & 91          & 24          & 9          & 176         \\ \hline
microsoft/vscode      & bug            & 87          & 20          & 13          & 180         \\
microsoft/vscode      & feature        & 80          & 14          & 20          & 186         \\
microsoft/vscode      & question       & 79          & 20          & 21          & 180         \\ \hline
bitcoin/bitcoin       & bug            & 89          & 18          & 11          & 182         \\
bitcoin/bitcoin       & feature        & 80          & 29          & 20          & 171         \\
bitcoin/bitcoin       & question       & 62          & 22          & 38          & 178         \\ \hline
opencv/opencv         & bug            & 80          & 8          & 20          & 192         \\
opencv/opencv         & feature        & 86          & 32          & 14          & 168         \\
opencv/opencv         & question       & 81          & 13          & 19          & 187         \\ \hline

\hline
 \end{tabular}
 \end{center}
\end{table}

Our models could likely benefit from additional tuning or training data to improve classification, especially for 'question' labels.
Addressing the imbalance between FP and FN across different labels could help improve model performance. This might involve re-evaluating the features used for classification or introducing class weights during training.

When compared to the baseline model results provided by the NLBSE Tool Competition Department our model performed slightly better overall and on some specific repos. The baseline model utilized the \textit{all-mpnet-base-v2} sentence transformer developed by hugging faces as their base model and fine-tuned it with SetFitTrainer. The baseline model had an overall average f1-score of 82.7\% when we got 82.8\%. The average f1-score on the tensorflow, bitcoin, and opencv repos for the baseline model was 86.44\%, 75.55\%, and 81.73\% respectively. Meanwhile, our models got 87.08\%, 76.65\%, and 82.5\% average f1-score on the same repos respectively. 

\section{Conclusion}

In conclusion, this study represents a significant step in applying large language models, specifically OpenAI's gpt-3.5-turbo, to classify GitHub issue reports. By fine-tuning models on datasets from five distinct repositories, we demonstrated the feasibility and efficiency of this approach. Our methodology, focusing on data preprocessing and model fine-tuning, yielded an average F1-score of 82.8\%, barely surpassing the baseline model's effectiveness in classifying issues into 'bug,' 'feature,' or 'question' categories. This performance, however, varied across repositories, revealing the nuanced nature of issue report classification.

One of the key findings was the variability in performance based on the nature of the data in each repository. This underscores the need for tailored approaches when applying language models in different contexts. Furthermore, the study highlighted challenges in classifying 'question' labels due to their often ambiguous nature. This points to a broader issue in the standardization of labeling practices within the GitHub community.

\section*{Acknowledgments} 
This work is supported in part by the NAU computer science research team, as well as Dr. Isac Artzi of GCU.

%\section{Acknowledgements}
%(Omitted to blind review)

%\clearpage

\bibliographystyle{ACM-Reference-Format}
\bibliography{references}

\end{document}